\documentstyle{mn}
\input{epsf.tex}
\newif\ifAMStwofonts

\ifoldfss
  \ifCUPmtlplainloaded \else
    \NewTextAlphabet{textbfit} {cmbxti10} {}
    \NewTextAlphabet{textbfss} {cmssbx10} {}
    \NewMathAlphabet{mathbfit} {cmbxti10} {} 
    \NewMathAlphabet{mathbfss} {cmssbx10} {} 
  \fi  %
  \ifAMStwofonts
    \ifCUPmtlplainloaded \else
      \NewSymbolFont{upmath} {eurm10}
      \NewSymbolFont{AMSa} {msam10}
      \NewMathSymbol{\upi}     {0}{upmath}{19}
      \NewMathSymbol{\umu}     {0}{upmath}{16}
      \NewMathSymbol{\upartial}{0}{upmath}{40}
      \NewMathSymbol{\leqslant}{3}{AMSa}{36}
      \NewMathSymbol{\geqslant}{3}{AMSa}{3E}

       \let\ge=\geqslant
    \fi
  \fi
\fi 

\ifnfssone
  \newmathalphabet{\mathit}
  \addtoversion{normal}{\mathit}{cmr}{m}{it}
  \addtoversion{bold}{\mathit}{cmr}{bx}{it}
  \newmathalphabet{\mathbfit} 
  \addtoversion{normal}{\mathbfit}{cmr}{bx}{it}
  \addtoversion{bold}{\mathbfit}{cmr}{bx}{it}
  \newmathalphabet{\mathbfss} 
  \addtoversion{normal}{\mathbfss}{cmss}{bx}{n}
  \addtoversion{bold}{\mathbfss}{cmss}{bx}{n}
  \ifAMStwofonts
    \ifCUPmtlplainloaded \else
      \UseAMStwoboldmath
      \makeatletter
      \new@mathgroup\upmath@group
      \define@mathgroup\mv@normal\upmath@group{eur}{m}{n}
      \define@mathgroup\mv@bold\upmath@group{eur}{b}{n}
      \edef\UPM{\hexnumber\upmath@group}
      \new@mathgroup\amsa@group
      \define@mathgroup\mv@normal\amsa@group{msa}{m}{n}
      \define@mathgroup\mv@bold\amsa@group{msa}{m}{n}
      \edef\AMSa{\hexnumber\amsa@group}
      \makeatother
      \mathchardef\upi="0\UPM19
      \mathchardef\umu="0\UPM16
      \mathchardef\upartial="0\UPM40
      \mathchardef\leqslant="3\AMSa36
      \mathchardef\geqslant="3\AMSa3E

       \let\ge=\geqslant
    \fi
  \fi
\fi 

\ifnfsstwo
  \DeclareMathAlphabet{\mathbfit}{OT1}{cmr}{bx}{it}
  \SetMathAlphabet\mathbfit{bold}{OT1}{cmr}{bx}{it}
  \DeclareMathAlphabet{\mathbfss}{OT1}{cmss}{bx}{n}
  \SetMathAlphabet\mathbfss{bold}{OT1}{cmss}{bx}{n}
  \ifAMStwofonts
    \ifCUPmtlplainloaded \else
      \DeclareSymbolFont{UPM}{U}{eur}{m}{n}
      \SetSymbolFont{UPM}{bold}{U}{eur}{b}{n}
      \DeclareSymbolFont{AMSa}{U}{msa}{m}{n}
      \DeclareMathSymbol{\upi}{0}{UPM}{"19}
      \DeclareMathSymbol{\umu}{0}{UPM}{"16}
      \DeclareMathSymbol{\upartial}{0}{UPM}{"40}
      \DeclareMathSymbol{\leqslant}{3}{AMSa}{"36}
      \DeclareMathSymbol{\geqslant}{3}{AMSa}{"3E}

       \let\ge=\geqslant
    \fi
  \fi
\fi 

\ifCUPmtlplainloaded \else
  \ifAMStwofonts \else 
    \def\upi{\pi}
    \def\umu{\mu}
    \def\upartial{\partial}
  \fi
\fi

\title{Transient chaos and resonant phase mixing in violent relaxation}

\author[H. E. Kandrup, I. M. Vass, and I. V. Sideris]
  {Henry E. Kandrup,$^{1,2,3}$\thanks{E-mail: kandrup@astro.ufl.edu}
    Ileana M. Vass$^{1}$\thanks{E-mail: vass@astro.ufl.edu}
    and Ioannis V. Sideris$^{1,4}$\thanks{E-mail: sideris@nicadd.niu.edu}\\
   $^{1}$ Department of Astronomy, University of Florida, Gainesville, 
          FL 32611, USA\\
   $^{2}$ Department of Physics, University of Florida, Gainesville, 
          FL 32611, USA\\
   $^{3}$ Institute for Fundamental Theory, University of Florida, Gainesville,
          FL 32611, USA\\
   $^{4}$ Department of Physics, Northern Illinois University, De Kalb, IL
	  60115 USA}

\date{Accepted 2002 \hskip 1in .
      Received 2002 \hskip 1in .}

\pagerange{\pageref{firstpage}--\pageref{lastpage}}
\pubyear{2000}

\begin{document}

\maketitle

\label{firstpage}

\begin{abstract}
This paper explores how orbits in a galactic potential can be impacted by 
large amplitude time-dependences of the form that one might associate with 
galaxy or halo formation or strong encounters between pairs of galaxies. A 
period of time-dependence with a strong, possibly damped, oscillatory 
component can give rise to large amounts of {\em transient chaos}, and it is
argued that chaotic phase mixing associated with this transient chaos could
play a major role in accounting for the speed and efficiency of violent
relaxation. Analysis of simple toy models involving time-dependent 
perturbations of an integrable Plummer potential indicates that this 
chaos results from a broad, possibly generic, resonance between the 
frequencies of the orbits and harmonics thereof and the frequencies of the 
time-dependent 
perturbation. Numerical computations of orbits in potentials exhibiting damped 
oscillations suggest that, within a period of $10$ dynamical times $t_{D}$ or 
so, one could achieve simultaneously both `near-complete' chaotic phase mixing
and a nearly time-independent, integrable end state.
\end{abstract}

\begin{keywords}
chaos -- galaxies: kinematics and dynamics -- galaxies: formation
\end{keywords}

\section{MOTIVATION}
Dating back at least to the pioneering work of Lynden-Bell (1967), it has 
been recognised that violent relaxation, viewed broadly as the mechanism 
whereby, even in the absence of dissipation, a galaxy evolves towards an
equilibrium or near-equilibrium state, must be a collective process
somehow involving phase mixing. However, not just any phase mixing will 
do: the simplest time-independent models which one might envision, such as 
Lynden-Bell's
balls rolling in a pig-trough, do not approach an equilibrium nearly fast
enough ({\em cf.} Kandrup 2001). For this reason, it has often been assumed 
that the fact that the potential is strongly time-dependent is crucial in 
making the mixing sufficiently efficient.

In the mid-1990's, however, it was recognised that it may not be time
dependence {\em per se} that is crucial. Rather, what seems essential is
that the flow be chaotic, exhibiting an exponentially sensitive dependence
on initial conditions. The important point here ({\em cf.} Kandrup and Mahon
1994, Mahon, Abernathy, Bradley, and Kandrup 1995, Merritt and Valluri 1996)
is that, whereas the phase mixing of regular orbits proceeds
as a modest power law in time, the phase mixing of chaotic orbits proceeds
exponentially. Thus, {\em e.g.,} a localised ensemble of initial conditions
corresponding to chaotic orbits will, when evolved in a time-independent 
Hamiltonian, begin by diverging exponentially from its localised state and 
then converge exponentially
towards an equilibrium or near-equilibrium state (Kandrup 1998). 

Unfortunately, however, most work on chaotic phase mixing hitherto has 
focused on flows in time-independent Hamiltonian systems, possibly allowing
for low amplitude perturbations ({\em cf.} Kandrup, Pogorelov, and Sideris
2000) but ignoring the possibility of a large amplitude time-dependence.
This work on (nearly) time-independent systems confirms that chaotic phase
mixing can be very efficient. However, it is clear that this phenomenon can 
be important only if the potential admits large measures of chaotic orbits.
The obvious point, then, is that time-independent density distributions
corresponding to galactic equilibria or near-equilibria are not likely to
contain huge measures of chaos. As stressed, {\em e.g.,} by Binney (1978), 
interestingly shaped equilibria which might be expected to admit global
stochasticity must contain large measures of regular orbits to serve as
the `skeleton' of the interesting structure. 

The obvious questions, then are: might one expect to have much larger measures
of chaos in realistic time-dependent systems? and should these orbits 
exhibit the qualitative
behaviour which one would expect to observe during an epoch of violent
relaxation? The aim of this paper is to argue that the answers to these 
questions is: Yes!

The crucially important point is that allowing for a time-dependent 
Hamiltonian can 
lead to resonances that trigger an epoch of {\em transient chaos}: even if the
initial and final density distributions are nearly integrable and admit no
global stochasticity, the intervening states could still correspond to
orbits exhibiting significant exponential sensitivity.

As a paradigmatic example, one can consider a test particle moving in a
spherically symmetric, constant density mass distribution which is, in
addition, subjected to a periodic sinusoidal perturbation corresponding
to a force per unit mass ${\bf F}=-{\beta}{\bf r}\cos {\omega}t$.
In the absence of the perturbation, a particle will feel an isotropic 
harmonic oscillator potential and, as such, the separation ${\delta}{\bf r}$ 
between two nearby orbits will execute simple harmonic motion.
However, the presence of the periodic driving complicates matters, leading
to a more complex equation for ${\delta}{\bf r}$, a typical component of which
will satisfy 
\begin{equation}
{d^{2}{\delta}r\over dt^{2}}+{\Omega}^{2}{\delta}r=0, \qquad {\rm with} \qquad
{\Omega}^{2}={\alpha}+{\beta}\cos{\omega}t,
\end{equation}
a classic example of a Hill equation. The obvious point, then, is that
even if ${\alpha}$ is positive and $|{\beta}|<{\alpha}$, ${\delta}r$ can
grow exponentially in time, corresponding to a chaotic orbit. The
periodic driving has triggered a parametric instability. 

Real systems are not constant density, so that the unperturbed 
(time-independent) orbits 
are more complex, and the time-dependent perturbations to which the 
system is subjected may not be strictly sinusoidal. However, as will be shown
in this paper, one might still expect generically that a system exhibiting
damped oscillations of the form associated with many scenaria involving
collective relaxation will exhibit a large amount of chaos.

That time-dependent perturbations can lead to transient chaos has been
recognised already in other branches of physics, including nonneutral 
plasmas ({\em cf.} Qian, Davidson, and Chen 1999, Strasburg and Davidson 2000) 
and charged particle beams (Gluckstern 1994). Indeed, Gluckstern has suggested
that chaos triggered by time-dependent perturbations may be responsible
for undesirable broadening (`emittance growth') in a focused charged particle 
beam by ejecting particles from the central core into an outerlying halo.

Section 2 of this paper begins by discussing some  `natural' physical 
expectations one might have regarding the possibility of transient chaos 
induced by time-dependent perturbations, and then describes a set of numerical 
experiments that were performed to test these expectations. Section 3
focuses on the possibility of transient chaos in its purest form, considering 
in detail what can happen when an integrable spherical system is subjected to
an undamped oscillatory perturbation  with a single frequency.
Section 4 extrapolates from this example to 
consider more realistic computations that capture two important aspects 
associated with violent relaxation: (1) {\em Allowing for damped oscillations.}
Can one trigger efficient chaotic mixing within (say) a period of $10$ 
dynamical times $t_{D}$ or so while simultaneously damping the original 
strongly time-dependent potential
to a nearly time-independent state? (2) {\em Allowing for a variable pulsation
frequency.} As the density distribution changes, one might expect that the
pulsation frequency will also change, and the obvious question is: do
such changes matter significantly? Section 5 summarises the principal
conclusions and describes extensions of the work described herein currently
underway.
\section{WHAT WAS EXPECTED AND WHAT WAS DONE}
\subsection{Physical expectations}

As stated already, the basic idea is that an appropriate time-dependence 
added to a potential can trigger chaos via a parametric resonance: if orbits
are subjected to a perturbation with power at appropriate frequencies, one
might expect otherwise regular orbits to become chaotic. In particular, if
unperturbed orbits have power concentrated at frequencies ${\sim}{\;}{\Omega}$ 
and the perturbation has significant power at frequencies ${\omega}$ 
sufficiently close to ${\Omega}$, there is the possibility of resonance
overlap which, as is well known ({\em cf.} Lichtenberg and Lieberman 1992),
can lead to the onset of global stochasticity.

As a particularly simple example, one knows that if a regular orbit with 
substantial power at some frequency ${\Omega}$ is perturbed by a sinusoidal 
perturbation with frequency ${\omega}={\Omega}$, the orbit will typically
exhibit a drastic response which can manifest sensitive dependence on initial
conditions. This is, {\em e.g.,} the physics responsible for resonant circuits
in elementary electronics. The real issue, one might argue, is simply: `how
close' must ${\Omega}$ and ${\omega}$ be to trigger strong exponential
sensitivity?

Detailed analysis of the effects of resonance overlap suggests generically
that the width of such resonances will be an increasing function of amplitude.
Even if a slightly `detuned' small amplitude perturbation has 
no apparent effect, a large perturbation with identical frequency may elicit 
a huge response. Indeed, the numerical models described in Sections 3 and 4 
of this paper indicate that, for large amplitude perturbations -- fractional 
amplitude of order $10-20\%$ or more -- the resonance can be very broad. In 
particular, these models reveal that, if the power in the orbits is 
concentrated at frequencies ${\sim}{\;}{\Omega}$, one can get a nontrivial 
increase in chaos -- in both the relative number of chaotic orbits and the 
value of the largest Lyapunov exponent -- for sinusoidal perturbations with
$0.1{\Omega}{\;}{\la}{\;}{\omega}{\;}{\la}{\;}30{\Omega}$.
In other words, the resonance can be more than two orders of magnitude wide!

The important point, then, is that if resonances of this sort are really so
broad, one might expect transient chaos to be extremely common, if not
ubiquitous, in time-dependent galactic potentials. Numerical simulations of
galaxy encounters and mergers and most models of galaxy and halo formation 
imply that a system approaching equilibrium will exhibit damped oscillations.
To the extent, however, that one is dealing with collisionless
relaxation, there is dimensionally only one natural time scale in the problem,
namely the dynamical $t_{D}{\;}{\sim}{\;}1/\sqrt{G{\rho}}$, with ${\rho}$ a
characteristic density. In particular, $t_{D}$ should set the oscillation time 
scale as well as the orbital time scale. The exact numerical values of these 
time scales will involve numerical coefficients which will in general be 
unequal and vary as a function of location within the galaxy. If, however,
one only needs to assume that the oscillation and orbital time scale agree to 
within an order of magnitude or so, it would seem likely that the conditions
appropriate for resonance-induced transient chaos could arise in much, if
not all, of the galaxy.

The crucial recognition, then, is that even a relatively short period of
exponential sensitivity can result in comparatively efficient chaotic phase
mixing. Earlier analysis of flows in time-independent potentials ({\em cf.}
Mahon, Abernathy, Bradley, \& Kandrup 1995, Merritt \& Valluri 1996, 
Kandrup 1998) have demonstrated that the presence of chaos can dramatically
enhance the degree to which a system becomes `shuffled' and, hence, the 
rate at which a localised orbit ensemble evolves towards an equilbrium. By
complete analogy, one might expect that transient chaos in a self-consistently
determined time-dependent potential will again enhance the efficacy of 
`mixing' and, by so doing, facilitate a more rapid approach towards an 
equilibrium.

It should be noted explicitly that the physical mechanism for the
generation of chaos described here is very different from the mechanism
whereby the amount of chaos can increase -- or decrease -- in more slowly
varying potentials ({\em cf.} Kandrup and Drury 1998, Contopoulos 2002).
If the potential is slowly varying, one can visualise an orbit as moving in
a phase space which is slowly changing but which is nearly constant over a
time scale ${\sim}{\;}t_{D}$. It is then natural to suppose that at some
times the orbit finds itself in a phase space region that is `chaotic' in the 
sense
that there is a sensitive dependence on initial conditions, but that at other
times it finds itself in `regular' regions where there is no such sensitive 
dependence. In the setting envisioned in this paper, the phase space is 
changing on a time scale comparable to -- or even shorter than -- the orbital
time scale, so that this picture is necessarily lost. Moreover, in the setting
described here one would expect generically a systematic {\em increase} in 
the amount and degree of chaos, not simply a change which could be either
positive or negative.

Finally, it should be noted that, to a certain extent, the term {\em transient
chaos} is necessarily `fuzzy.' Chaos, like ordinary Lyapunov exponents, is
only defined in an asymptotic $t\to\infty$ limit. The point, however, is that,
even over finite time intervals, it is physically meaningful to ask whether 
orbits exhibit an exponentially sensitive dependence on initial conditions.
This is, {\em e.g.,} the notion that motivates the definition of finite time
Lyapunov exponents ({\em cf.} Grassberger, Badii, and Poloti 1988), which
have become accepted tools in nonlinear dynamics. From a phenomenological
point of view, it makes perfect sense to identify an orbit as exhibiting
significant transient chaos if, for some finite time $>t_{D}$, an orbit
exhibits significant exponential sensitivity. In many cases, it is clear
by visual inspection and/or through a computation of a finite time Lyapunov
exponent whether such transient chaos is present. In other cases, however,
things are not so completely clear. For example, for the models considered
in this paper it proves difficult to determine with complete reliability 
a minimum frequency at which transient chaos `turns on' since this `turning
on' is comparatively gradual.
\subsection{Numerical experiments}
The numerical computations described in this paper involved perturbations
of a Plummer sphere, which is characterised (in units with $G=m=1$) by the
potential
\begin{equation}
V_{0}(x,y,z)=-{1\over (1+x^{2}+y^{2}+z^{2})^{1/2}}.
\end{equation}
This form was selected (1) for its inate simplicity and, especially, (2) 
because orbits in this potential are all strictly integrable. Any chaos that
is observed can be associated unambiguously with the perturbations that were
introduced. These were assumed to take the form
\begin{equation}
V_{1}(x,y,z,t)=-{m\over (1+x^{2}+a^{2}y^{2}+z^{2})^{1/2}},
\end{equation}
where the quantities $m$ and $a$ are allowed to vary in time. A variety of
different time-dependences were considered, including:
\par\noindent 
${\bullet}$ strictly sinusoidal oscillations in $m$ or $a$, {\em i.e.,}
\begin{equation}
m=m_{0}\sin{\omega}t \qquad {\rm and} \qquad
a=a_{0}+{\delta}a\sin{\omega}t.
\end{equation}
This allowed one to focus on the physical mechanism in its `purest' form.
\par\noindent 
${\bullet}$ sinusoidal oscillations that damp exponentially or as a power 
law in time, {\em i.e.,}
\begin{equation}
m=m_{0}e^{-{\alpha}t}\,\sin {\omega}t
\end{equation}
or
\begin{equation}
m=m_{0}{\sin {\omega}t\over (t_{0}+t)^{p}},
\end{equation}
with $p=1$ or $2$, and 
analogous formulae for $a(t)$. This allowed one to confirm that the physical
effects associated with a purely oscillatory perturbation persist if, as is
natural in the context of violent relaxation, one allows for a galaxy that
damps towards equilibrium.
\par\noindent
${\bullet}$ nonoscillatory damping, {\em i.e.,} 
\begin{equation}
m=m_{0}e^{-{\alpha}t}
\end{equation}
or
\begin{equation}
m={m_{0}\over (t_{0}+t)^{p}}
\end{equation}
and analogous formulae for $a(t)$. This allowed one to demonstrate that the
presence of an oscillatory component is crucial in order to get a substantial
amount of transient chaos. As will be described below, computations with
perturbations given by eq.~(7) or (8) typically exhibited comparatively
minimal amounts of transient chaos.
\par\noindent ${\bullet}$ variable driving frequency, assuming that
\begin{equation}
{\omega}(t)={\omega}_{0}+{\delta}(t),
\end{equation}
with ${\delta}(t)$ a randomly varying variable that samples an 
Ornstein-Uhlenbeck process ({\em cf.} Van Kampen 1981). What this means 
is that ${\delta}$ is treated
as Gaussian coloured noise, so that its statistical properties are determined
uniquely by its first two moments, which are assumed to satisfy
\begin{equation}
{\langle}{\delta}(t){\rangle}=0 {\rm\;\;\; and \;\;\;}
{\langle}{\delta}(t_{1}){\delta}(t_{2}){\rangle}={\Delta}^{2}\exp(-
|t_{1}-t_{2}|/t_{c}).
\end{equation}
Here ${\Delta}$ represents the typical `size' of the random component and
$t_{c}$ the time scale over which it changes appreciably. Considering
perturbations of this form allowed one to confirm that, because the resonance
is comparatively broad, permitting the driving frequency to drift (within
limits) has a relatively minor -- albeit potentially significant -- effect. 
This again is important physically.
Since the density distribution varies as a galaxy evolves towards an
equilibrium, one would anticipate that the characteristic oscillation 
frequency (or frequencies) will change.

Computations were performed for ensembles of 1600 initial conditions which
were generated in two different ways. (1) Broad `representative' ensembles 
were generated by uniformly sampling a constant energy hypersurface. A study
of orbits generated from such ensembles allowed one to derive systematic 
effects which might be expected to act in a galaxy as a whole. (2) Localised 
ensembles were generated by sampling small phase space hypercubes. Studying
orbits generated from such ensembles allowed one to determine the extent to 
which the `smooth' behaviour associated with the representative ensembles 
reflected the choice of ensemble as opposed to properties of the resonance.

Most experiments involved perturbations at the $50\%$ level or less, {\em
i.e.,} $m_{0}{\;}{\la}{\;}0.5$ and/or ${\delta}a{\;}{\la}{\;}0.5$. For the
intermediate energies considered throughout, a `typical' orbital frequency
(derived from a Fourier transform) was ${\Omega}{\;}{\sim}{\;}0.3-0.35$, 
corresponding to $t_{D}=2{\pi}/{\Omega}{\;}{\sim}{\;}20$ in
absolute units. On physical grounds one might expect that the perturbations 
should damp on a time scale comparable to, but somewhat longer than, $t_{D}$,
so that most experiments assumed that (to within factors of a few) 
$t_{0}$ and ${\alpha}^{-1}{\;}{\sim}{\;}
5t_{D}{\;}{\sim}{\;}100$. (Computations of chaotic phase mixing in 
time-independent potentials [{\em cf.} Kandrup \& Novotny 2002] typically 
exhibit an exponential approach towards
a near-invariant distribution on a time scale ${\sim}{\;}5\,t_{D}$!)

\section{PARAMETRIC RESONANCE AND TRANSIENT CHAOS}
Computations involving representative orbit ensembles have led to several
unambiguous conclusions:
\par\noindent
1. Integrations which do not involve large scale time variations on a time
scale ${\sim}{\;}t_{D}$ yield little, if any, transient chaos.
In particular, as will be illustrated more carefully in the following Section,
allowing for nonoscillatory damped perturbations of the form given by eq.~(7)
do not result in large measures of chaos, even for large amplitudes 
$m_{0}{\;}{\ga}{\;}0.5$. However, allowing for oscillations in $m$ and/or
$a$ {\em can} trigger substantial amounts of chaos. 
\par\noindent
2. There are at least two solid reasons to believe that this chaos is triggered
by a resonance. More obvious, perhaps, is the fact that the largest
increases in chaos -- as reflected by both the number of chaotic orbits 
and the size of a typical Lyapunov exponent -- arise for driving frequencies
comparable to, or somewhat larger than, the frequencies for which the 
unperturbed orbits have most of their power. Equally significant, however,
is the fact that the driving frequencies which result in the greatest amount
of chaos correspond precisely to those frequencies for which the orbital
energies are `shuffled' the most, {\em i.e.,} where different particles 
with the same initial energy end up with the largest spread in final energies. 
{\em A priori} there need be no direct 
connection between increases in chaos and large-scale shuffling of energies. 
However, one {\em would} expect resonant couplings to lead to significant 
changes in energies. A correlation between changes in energy and the amount 
of chaos thus corroborates the interpretation that this chaos is resonant 
in origin.
\par\noindent
3. At least for the case of large amplitude oscillations, 
$m_{0}{\;}{\ga}{\;}0.1$ or so, the resonance is comparatively broad. 
One observes significant increases in both the relative
measure of chaotic orbits {\em and} the size of the largest Lyapunov exponent
whenever the driving frequency ${\omega}$ and the natural orbital frequencies
${\Omega}$ agree to within an order of magnitude or so. If, {\em e.g.,} one
considers an energy where the natural frequencies peak for 
${\Omega}_{m}{\;}{\sim}{\;}0.3-0.35$ or so, noticeable increases in the degree 
of chaos are observed for $0.035{\;}{\la}{\;}{\omega}{\;}{\la}{\;}10.0$,
{\em i.e.,} $0.1{\;}{\sim}{\;}{\omega}/{\Omega}_{m}{\;}{\sim}{\;}30$.
\par\noindent
4. The width of the resonance, {\em i.e.,} the range of values of ${\omega}$ 
for which one sees an appreciable increase in the degree of chaos, appears to 
be an increasing function of amplitude. 

\begin{figure}
\centering
\centerline{
    \epsfxsize=8cm
    \epsffile{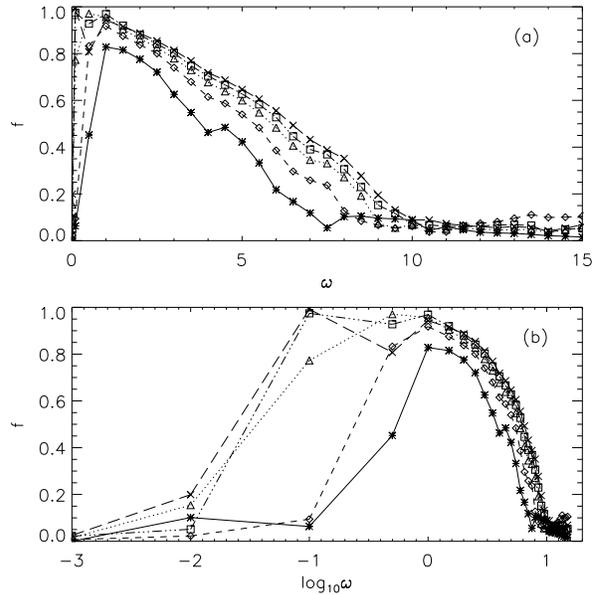}
      }
    \begin{minipage}{10cm}
    \end{minipage}
    \vskip -0.1in\hskip -0.0in
\caption{(a) The relative fraction $f$ of chaotic orbits in a representative 
1600 orbit ensemble which are subjected to strictly sinusoidal oscillations 
with variable frequency ${\omega}$. The different curves represent, from
bottom to top, driving
amplitudes $m_{0}=0.1$, $0.2$, $0.3$, $0.4$, and $0.5$. (b) The same data
plotted as a function of $\log_{10}{\omega}$.
}
\label{landfig}
\end{figure}

\begin{figure}
\centering
\centerline{
    \epsfxsize=8cm
    \epsffile{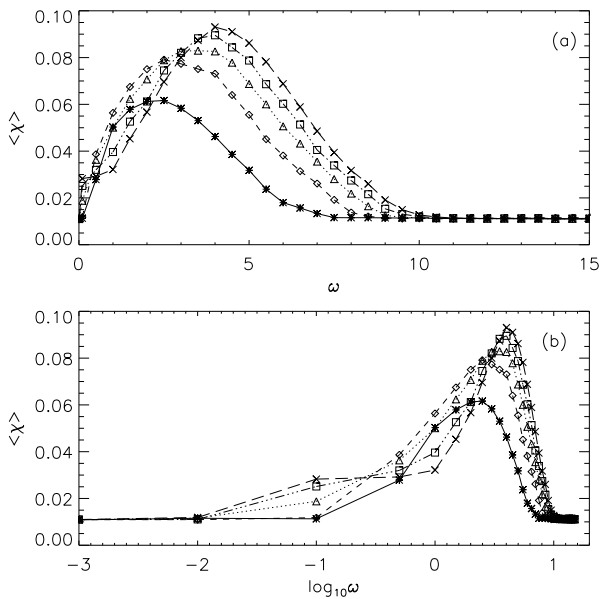}
      }
    \begin{minipage}{10cm}
    \end{minipage}
    \vskip -0.1in\hskip -0.0in
\caption{(a) The mean value ${\langle}{\chi}{\rangle}$ of the chaotic orbits
in the ensembles used to generate the preceding Figure, again plotted as a
function of ${\omega}$. 
(b) The same data plotted as a function of $\log_{10}{\omega}$.
}
\label{landfig}
\end{figure}

\par\noindent
5. The resonance appears to be smooth in the sense that, for fixed amplitude, 
the relative number of chaotic orbits and the size of the largest Lyapunov 
exponent both vary smoothly as functions of the driving frequency ${\omega}$.
In particular, plots of the fraction $f$ of the orbits that are chaotic
or the mean size ${\langle}{\chi}{\rangle}$ of the largest Lyapunov exponent 
as functions of driving frequency typically exhibit a single peak. There is
no obvious evidence for complex structure within the resonance. 
\par\noindent
6. For fixed ${\omega}$, the relative number of chaotic orbits and the size
of the largest Lyapunov exponent are both an increasing function of 
perturbation amplitude, although it appears that the fraction $f$ of orbits 
that are chaotic can asymptote towards a fixed value $f_{M}$ for sufficiently 
large amplitude. For some values of ${\omega}$, especially those near the 
center of the resonance, this $f_{M}$ can approach unity; for other choices 
of frequency, $f_{M}$ can be substantially smaller.
\par\noindent
7. Whether or not the perturbation breaks spherical symmetry appears to be
relatively unimportant, {\em e.g.,} making the parameter $a$ time-dependent 
or otherwise different from unity does not change things all that much. 
Provided that the system manifests large amplitude oscillations, one can 
apparently get just as much chaos for spherical systems as for
nonspherical systems.

Evidence for several of these conclusions is provided in Figures 1 - 4, all
generated from the same set of 1600 initial conditions, evolved for a total
time $t=512$ in the presence of a strictly sinusoidal perturbation of the 
form given by eq.~(4) with $a{\;}{\equiv}{\;}1$. 
The two panels of Figure 1 exhibit the relative measure $f$ of chaotic orbits 
as a function of driving frequency ${\omega}$ for five different 
choices of amplitude $m_{0}$. The top panel exhibits $f({\omega})$ on a
linear scale, thus allowing one to focus on the behaviour of the resonance
at higher frequencies; the lower exhibits $f$ as a function of 
$\log_{10}{\omega}$, which allows one to see more clearly the behaviour
at lower frequencies. 

Most obvious, perhaps, from Figure 1 is the fact that the width 
of the resonance is an increasing function of $m_{0}$, although this increase
appears to be comparatively small for $m_{0}>0.2$ or so. Also evident is the 
fact that, except perhaps for the lowest amplitude, $m_{0}=0.1$, the fraction 
$f$ appears to vary smoothly with frequency. In each case, $f$ peaks at a 
frequency ${\omega}_{max}$ comparable to the typical natural frequencies 
${\Omega}{\;}{\sim}{\;}0.3-0.35$ associated with the unperturbed orbits.
However, ${\omega}_{max}$ {\em does} seem to vary somewhat as a function of 
$m_{0}$. For larger amplitudes, near ${\omega}_{max}$ essentially all the 
orbits are chaotic, {\em i.e.,} $f({\omega}_{max}){\;}{\approx}{\;}1$. For 
lower frequencies $f({\omega}_{max})$ can be significantly less 
than unity. Overall, there is evidence for significant amounts
of transient chaos for $0.35{\;}{\la}{\;}{\omega}{\;}{\;}{\la}{\;}10.0$.

The two panels of Figure 2 exhibit ${\langle}{\chi}{\rangle}$, the mean 
value of the largest finite time Lyapunov exponent for the same ensembles, 
again plotted on both linear and logarithmic scales. Here 
${\langle}{\chi}{\rangle}$ was extracted by first identifying those orbits 
in the ensemble that were deemed chaotic and then computing the mean value 
of ${\chi}$ for those orbits. 

Figure 3 exhibits ${\delta}E$, the root mean spread in energies at time
$t=512$ for the chaotic orbits identified in Figures 1 and 2. It is clear 
visually that, as one would expect if the transient chaos is associated with 
a resonant coupling, the frequencies which result in the largest measures
of chaotic orbits and the largest ${\langle}{\chi}{\rangle}$ also result
in the largest shifts in energy. However, it is also evident that, especially
for the higher amplitude perturbations, ${\delta}E$ exhibits a more
complex dependence on ${\omega}$ than do $f({\omega})$ or 
${\langle}{\chi}({\omega}){\rangle}$. This does not contradict the assertion
that transient chaos is induced by a resonant coupling, but it {\em does}
suggest that the amount and degree of chaos in the broad resonance region
is less sensitive to the pulsation frequency than is the shuffling in
energies. This has potentially significant implications for violent relaxation,
where one is interested in both (i) shuffling the energies of the individual
masses and (ii) randomising their phase space locations on a constant energy
hypersurface.

One other point is also evident from this Figure, namely that, at least for 
frequencies well into the resonance, the spread in energies can be very large. 
This reflects the fact that, because of the resonant coupling, a significant 
fraction of the orbits have acquired large energies that set them along 
trajectories involving excursions to very large radii, $r>10$ or more. 
Allowing for a resonance that is this strong to last for as long as 
$t{\;}{\approx}{\;}20t_{D}$ is completely unrealistic in the context of any 
model for violent relaxation. The computations described in Section 4b, which 
incorporate strong damping, avoid this artificial behaviour. 

In part because of this effect, any uniform prescription used to distinguish 
between `regular' and `chaotic' orbits is necessarily somewhat {\em ad hoc}
and, as such, fraught with difficulties. Figs.~1 and 2 reflect an analysis
in which `chaotic' was defined as corresponding to a finite time Lyapunov 
exponent for the interval $0<t<512$ with a value ${\chi}{\;}{\ge}{\;}0.012$,
arguably a fairly large threshhold value. Lowering the threshhold value of 
$0.012$ to a smaller value would of course increase the estimated fraction $f$ 
of chaotic orbits and decrease the mean value of ${\langle}{\chi}{\rangle}$ 
for those orbits deemed chaotic. In any event, the choice 
${\chi}{\;}{\ge}{\;}0.012$ implies that there is little if any chaos for 
${\omega}\;{\la}\;0.01$ and ${\omega}\;{\ga}\;10$, which is consistent with 
the fact that localised ensembles exhibit little if any evidence of chaotic 
phase mixing for ${\omega}<0.01$ and 
${\omega}>10$ or so. Despite this, however, it should be recognised explicitly 
that this prescription really identifies orbits which are `obviously' or 
`strongly' chaotic, and that it may be ignoring a considerable number of 
orbits which are only very weakly chaotic.

This prescription, like any other which involves a fixed 
minimum value and a fixed integration time, is open to criticism. For 
those frequencies and amplitudes where the resonance is especially strong and 
many orbits quickly achieve large energies, it might seem more appropriate to 
consider a shorter time interval, during which the orbits are restricted to 
smaller radii. However, for frequencies and amplitudes for which the resonance 
is weaker, one might wish instead to integrate for much longer times so as 
(hopefully) to allow for clear distinctions to be made between regular and 
`extremely sticky', nearly regular ({\em cf.} Contopoulos 1971) orbits. The 
choice of a time interval $t=512$ represents a compromise necessitated by the 
idealised nature of the computations described in this section.

In any event, what is inevitable is that, for chaotic orbits in any 
potential, one must expect correlations between the energy
of the orbit and the size of a typical finite time Lyapunov exponent
${\chi}$. For the potential 
considered here, one finds typically that the orbits deemed regular are 
precisely those which exhibit the smallest changes in energy and hence, 
overall, the smallest excursions from the origin. However, for those 
orbits which {\em are} chaotic, those spending more time at larger radii
and which, for this reason, have larger orbital time scales, tend to have 
somewhat smaller values of ${\chi}$.

\begin{figure}
\centering
\centerline{
    \epsfxsize=8cm
    \epsffile{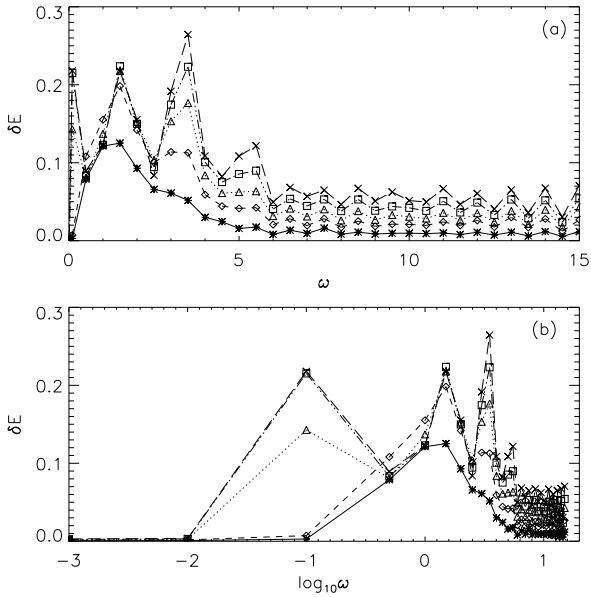}
      }
    \begin{minipage}{10cm}
    \end{minipage}
    \vskip -0.1in\hskip -0.0in
\caption{(a) The root mean squared spread in energies,
${\delta}E_{rms}(t=512)$, for the orbit ensembles used to generate the 
first two Figures. 
(b) The same data plotted as a function of $\log_{10}{\omega}$.
}
\label{landfig}
\end{figure}

\begin{figure}
\centering
\centerline{
    \epsfxsize=8cm
    \epsffile{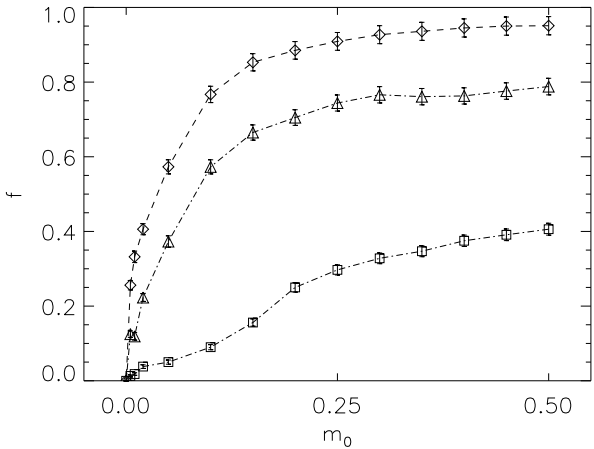}
      }
    \begin{minipage}{10cm}
    \end{minipage}
    \vskip -0.1in\hskip -0.0in
\caption{
The relative fraction $f$ of chaotic orbits in a representative 1600 orbit
ensemble which are subjected to strictly sinusoidal oscillations of the
form given by eq.~(4) of variable amplitude $m_{0}$ with
${\omega}=1.4$ (upper curve), ${\omega}=3.5$ (middle curve), and
${\omega}=7.0$ (lower curve).
}
\label{landfig}
\end{figure}

Figure 4 exhibits the relative measure of chaotic orbits as a function of
amplitude $m_{0}$ for three different frequencies, namely
${\omega}=1.4$, $3.5$, and $7.0$. For ${\omega}=1.4$, a frequency
in the middle of the resonance, for amplitudes as large as $m_{0}=0.25$
almost all the orbits exhibit evidence of chaos. For ${\omega}=3.5$, a value
somewhat closer to the edge of the resonance, the relative abundance of 
chaotic orbits is somewhat smaller. However, for $m_{0}=0.25$ as many as 
$75\%$ of the orbits are clearly chaotic, a fraction which does not increase
significantly if $m_{0}$ is increased. For ${\omega}=7.0$, a value 
near the edge of the resonance, the fraction of the orbits which is chaotic
is smaller and does {\em not} appear to level off for $m_{0}>0.25$ or
so. Indeed, if the amplitude is increased to a value as large as $m_{0}=1.0$
as many as $50\%$ of the orbits prove chaotic, {\em i.e.,} $f(m_{0}=1)
{\;}{\approx}{\;}0.5$.

The uniform criterion used to identify chaotic orbits for Figures 1 and 2
fails for $m_{0}{\;}{\la}{\;}0.1$. For that reason, Figure 4 implemented a 
different,
somewhat more subjective criterion which has been used successfully in 
distinguishing between regular and `very sticky' chaotic orbits in the 
triaxial generalisations of the Dehnen potentials (Siopis \& Kandrup 2000). 
What this entailed was ordering the $1600$ computed values of ${\chi}$ at 
various times $T_{A}$, plotting the ordered list of ${\chi}(T_{A})$'s for 
different values of $T_{A}$, and then in each case searching for a `kink' 
in the curve. This prescription appears to permit an accurate determination 
of a fraction $f$ of the orbits exhibiting chaotic behaviour which is 
insensitive to the precise choice of $T_{A}$, although the estimated size of 
the finite time ${\chi}(T_{A})$ {\em does} depend on the sampling interval.

Additional insights into the nature of the resonance can be derived from
Figure 5, which shows the $x$-components of the composite (normalised) Fourier 
transforms of coordinate and force per unit mass, $|x({\Omega})|$ and 
$|a_{x}({\Omega})|$, constructed by combining spectra for 1600 unperturbed 
${\omega}=0$ orbits generated from the same initial conditions used to 
generate Figures 1 - 3. (Because of spherical symmetry the $y$- and 
$z$-components are statistically identical.) As noted already, for this
particular choice of energy,  $|x({\Omega})|$ peaks at 
${\Omega}_{m}{\;}{\sim}{\;}0.3-0.35$ and is comparatively negligible for much
higher and lower frequencies. By contrast, $|a_{x}({\Omega})|$ has 
multiple peaks associated with higher frequency harmonics which arise because 
the force per unit mass is a nonlinear function of position. 
It is natural to suppose that it is these harmonics that are responsible for
the comparatively large responses at frequencies ${\Omega}$ large compared
with ${\Omega}_{m}$ that are evident in Figures 1 - 3. It is particularly 
interesting that this response cuts off for driving frequencies ${\omega}$ 
large compared with $2{\Omega}$, rather than ${\Omega}$, a fact that would
suggest that one is observing the effects of a broad $2:1$ resonance similar
to that which arises in the standard Matthieu equation (see, {\em e.g.,}
Matthews and Walker 1964).

\begin{figure}
\centering
\centerline{
    \epsfxsize=8cm
    \epsffile{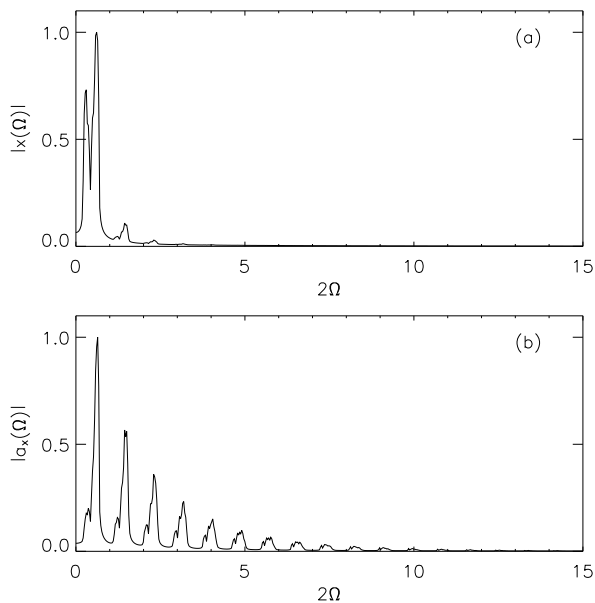}
      }
    \begin{minipage}{10cm}
    \end{minipage}
    \vskip -0.1in\hskip -0.0in
\caption{(a) The power spectrum $|x({\Omega})|$ associated with the 
$x$-component of the orbits 
generated from the 1600 initial conditions used to created Figs. 1 - 3, 
evolved in the absence of any time-dependent perturbation and plotted as
a function of $2{\Omega}$.
(b) The power spectrum $|a_{x}({\Omega})|$ associated with the $x$-component
of the force per unit mass associated with the same orbits.
}
\label{landfig}
\end{figure}

Much of the smoothness in plots such as those given in Figure 2 reflects the
fact that the data were generated from `representative' ensembles of initial
conditions. If the same analysis is repeated for different localised ensembles 
of initial conditions, even ensembles with the same energy, one observes
signficant variability in both the size of a typical finite time Lyapunov
exponent and the value of the frequency ${\omega}$ for which 
${\langle}{\chi}{\rangle}$ is maximised. And similarly, plots of quantities
like ${\langle}{\chi}{\rangle}$ as a function of ${\omega}$ for such
ensembles exhibit more
structure than do plots for representative ensembles. This presumably reflects
the fact that the natural frequencies of the otherwise regular orbits vary
with phase space location, and that some frequencies are more susceptible
to the resonance than others. This additional structure could have important
implications for near-equilibria subjected to nearly periodic perturbations,
which could impact certain orbits much more than others. 
However, one might expect that, when considering the `global' properties of
violent relaxation, such details would tend to wash out.

\par\noindent
\section{A TOY MODEL FOR VIOLENT RELAXATION}
\subsection{Variable pulsation frequencies}

Allowing for a randomly varying frequency of the form given by eqs.~(9) and
(10) leads to several significant conclusions.
\par\noindent
1. Perhaps the most important is that allowing for a comparatively slow 
frequency drift can actually {\em increase the relative measure of chaotic
orbits.} Suppose that the unperturbed frequency ${\omega}_{0}$ is well into 
the resonant region and that ${\Delta}$ is not sufficiently large to push 
${\omega}$ outside. In this case, the introduction of a perturbation 
${\delta}(t)$ with an autocorrelation time $t_{c}$ long compared with $t_{D}$ 
tends to convert many, if not all, the `regular' orbits into orbits with 
appreciable exponential sensitivity. This behaviour can be understood by 
supposing that, in the absence of the frequency drift, there exist some 
regular orbits which, because of the particular values of their natural 
frequencies, barely manage to avoid resonating with the perturbation. Allowing 
for a drift facilitates an improved resonance coupling which can make (some of)
these orbits chaotic for at least some of the time.
\par\noindent
2. Provided that ${\omega}_{0}$ and ${\omega}_{0}{\;}{\pm}{\;}{\Delta}$ are 
well within the resonance region, it is also apparant that, as the 
autocorrelation time $t_{c}$ decreases, the degree of chaos, as probed by the 
mean ${\langle}{\chi}{\rangle}$, tends to decrease, even though the 
relative fraction of chaotic orbits may well remain equal to unity. 
\par\noindent
3. For fixed autocorrelation time $t_{c}$, the mean ${\langle}{\chi}{\rangle}$ 
manifests only a relatively weak dependence on ${\Delta}$, at least for 
$0.1{\;}{\la}{\;}{\Delta}/{\omega}_{0}{\;}{\la}{\;}0.8$. For some choices
of $t_{c}$ increasing ${\Delta}$ causes an increase in the mean finite
time ${\chi}$; in other cases increasing ${\Delta}$ decreases the mean.
However, the observed variations are invariably small.

\begin{figure}
\centering
\centerline{
    \epsfxsize=8cm
    \epsffile{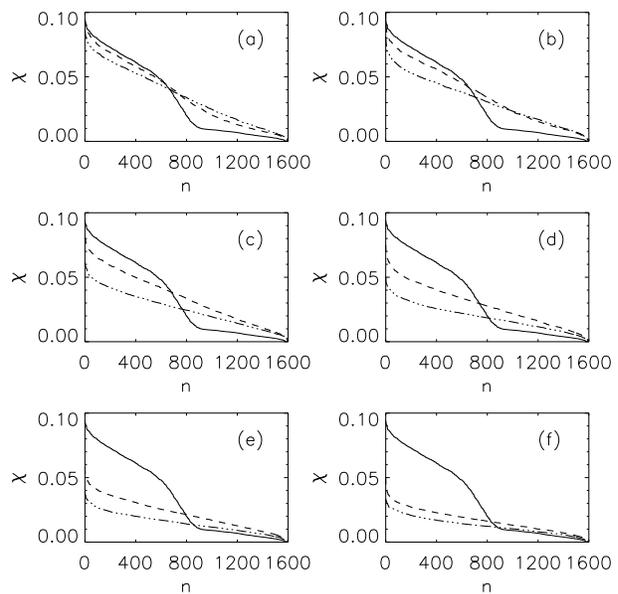}
      }
    \begin{minipage}{10cm}
    \end{minipage}
    \vskip -0.1in\hskip -0.0in
\caption{(a) Finite time Lyapunov exponents for 1600 representative orbits 
evolved for a time $t=512$ in an undamped Plummer potential pulsed with 
amplitude $m=0.1$ and a frequency
frequency ${\omega}(t)={\omega}_{0}+{\delta}(t)$, with ${\omega}_{0}=3.5$,
autocorrelation
time $t_{c}=320=16t_{D}$, and ${\Delta}^{2}=0.35$ (dashed curve) and 
${\Delta}^{2}=2.8$ (triple-dot-dashed). The solid curve contrasts exponents 
derived for an evolution with ${\delta}{\;}{\equiv}{\;}0$. (b) The same
for $t_{c}=160$. (c) $t_{c}=80$. (d) $t_{c}=40$. (e) $t_{c}=20$. 
(f) $t_{c}=10$.
}
\label{landfig}
\end{figure}

\begin{figure}
\centering
\centerline{
    \epsfxsize=8cm
    \epsffile{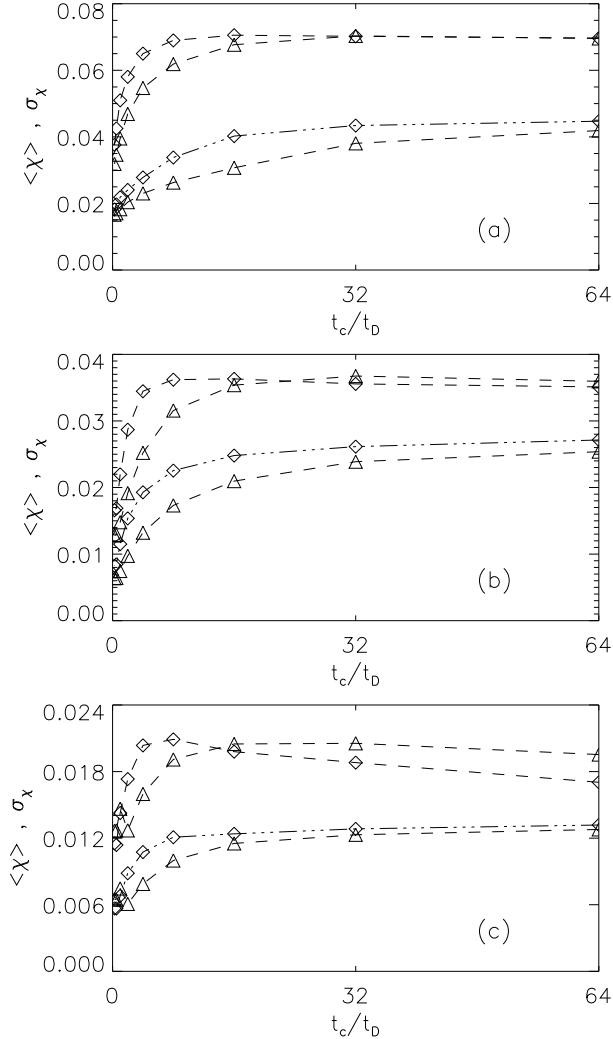}
      }
    \begin{minipage}{10cm}
    \end{minipage}
    \vskip -0.1in\hskip -0.0in
\caption{(a) The mean Lyapunov exponent ${\langle}{\chi}{\rangle}$ (upper two
curves) and the associated dispersion ${\sigma}_{\chi}$ (lower two curves) 
computed for ensembles of 1600 orbits evolved in an undamped Plummer potential
with amplitude $m=0.5$, ${\omega}_{0}=3.5$, and variable autocorrelation 
time $t_{c}$. Diamonds
correspond to ${\Delta}^{2}=0.35$, triangles to ${\Delta}^{2}=2.8$.
(b) The same for $m_{0}=0.1$. (c) $m_{0}=0.05$.
}
\label{landfig}
\end{figure}

All three of these points are illustrated in Figures 6 and 7. The dashed and
triple-dot-dashed curves in Figure 6 correspond to plots of ordered finite 
time Lyapunov exponents ${\chi}$ associated with $1600$ initial conditions 
subjected to undamped pulsations with variable frequency 
${\omega}={\omega}_{0}+{\delta}(t)$ for $m_{0}=0.1$, ${\omega}_{0}=3.5$ 
and different choices of ${\Delta}$ and $t_{c}$. The solid curve in each 
panel contrasts the exponents generated for the same initial conditions
evolved with ${\delta}=0$. It is evident that, for the largest values of 
$t_{c}$, the differences between the curves with ${\delta}$ zero and nonzero 
are relatively small, at least for the largest values of ${\chi}$. However, 
it is also apparent that the time-dependent frequency drift has significantly 
reduced the relative number of orbits with very small finite time ${\chi}$'s. 
Also evident is the fact that decreasing the autocorrelation time $t_{c}$,
{\em i.e.,} making the frequency drift more quickly, tends generically to
decrease the typical value of ${\chi}$, although the relative measure of 
chaotic orbits may not change appreciably.
For this particular choice of ${\omega}_{0}$ and $m_{0}$, increasing 
${\Delta}$ for fixed $t_{c}$ also tends to reduce the typical size of a finite
time ${\chi}$. This, however, is {\em not} generic. If, instead, the same 
computations are repeated for $m_{0}=0.05$, one finds that the choice of
${\Delta}$ which leads to the largest ${\chi}$'s actually varies with $t_{c}$.

The effects of varying ${\Delta}$ and $t_{c}$ on the mean 
${\langle}{\chi}{\rangle}$ and the dispersion
${\sigma}_{\chi}$ associated with these ensembles is exhibited in Figure 7.

\subsection{Damped oscillations}

Allowing for damped oscillations of the form given by eq.~(5) or (6) has
a comparatively minimal effect. Not surprisingly, the size of the mean 
Lyapunov exponent ${\langle}{\chi}{\rangle}$ decreases as time elapses and
there can be a gradual decrease in the relative fraction $f$ of orbits
exhibiting exponential sensitivity. However, the basic phenomenon of transient
chaos persists until the perturbation has damped almost completely away.
\begin{figure}
\centering
\centerline{
    \epsfxsize=8cm
    \epsffile{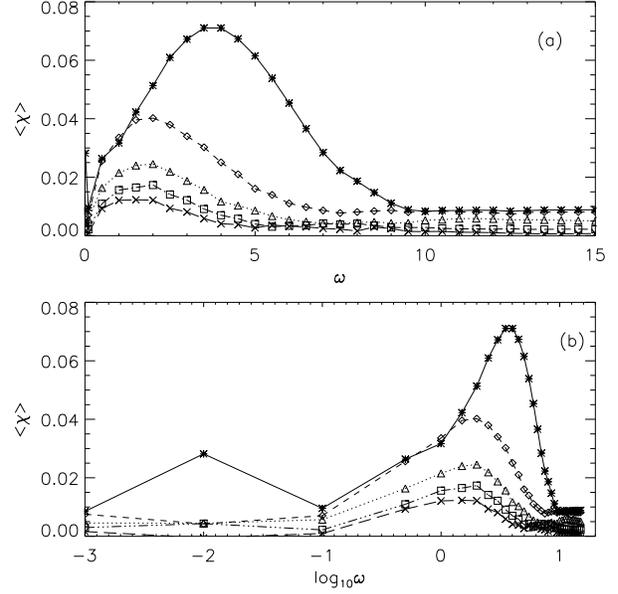}
      }
    \begin{minipage}{10cm}
    \end{minipage}
    \vskip -0.1in\hskip -0.0in
\caption{(a) The four lower curves exhibit the mean value 
${\langle}{\chi}{\rangle}$ as a function of ${\omega}$ for chaotic orbits 
subjected to damped oscillations with $m_{0}=0.5$, $p=2$, and $t_{0}$ for
the intervals (from top to bottom) $100<t<200$, $200<t<300$, $300<t<400$, and
$400<t<500$. The top curve exhibits ${\langle}{\chi}{\rangle}$ for undamped
($p=0$) oscillations for the intervals $0<t<512$.
(b) The same data plotted as a function of $\log_{10}{\omega}$.
}
\label{landfig}
\end{figure}

This is, {\em e.g.,} evident from Figure 8, which was again generated from
the same 1600 initial conditions used to generate Figures 1 - 5, now allowing
for a perturbation of the form given by eq.~(6) with $m_{0}=0.5$, $t_{0}=100$,
and $p=2$, again allowing for variable ${\omega}$. Here the four lower 
curves (from top to bottom) represent mean values of the finite time Lyapunov
exponent ${\langle}{\chi}{\rangle}$ generated separately for the intervals 
$100<t<200$, $200<t<300$, $300<t<400$, and $400<t<500$. The top curve,
reproduced from Fig.~2, represents ${\langle}{\chi}{\rangle}$ for $0<t<512$
for the same orbits evolved in the presence of undamped ($p=0$) oscillations
with the same initial $m_{0}$. It is evident that, as one would have expected, 
the degree of exponential sensitivity decreases with time and that, by the
end of the integration, the orbits are behaving in a nearly regular fashion.
The fact that the peak frequency ${\omega}$ drifts towards lower values
at later times is consistent with the fact, evident from Fig.~2, that for
undamped oscillations, lower amplitudes $m_{0}$ correlate with lower values
of ${\omega}$.

\begin{figure}
\centering
\centerline{
    \epsfxsize=8cm
    \epsffile{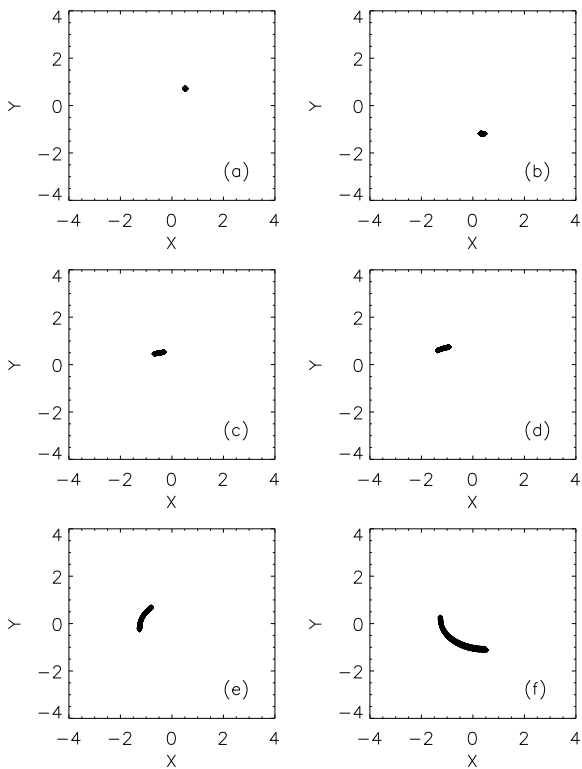}
      }
    \begin{minipage}{10cm}
    \end{minipage}
    \vskip -0.3in\hskip -0.0in
\caption{The $x$ and $y$ coordinates of an initially localised ensemble of 
orbits evolved in a Plummer potential subjected to a nonoscillatory 
perturbation of the form given by eq.~(7) with $m_{0}=0.5$ and $p=2$.
(a) $t=0$. (b) $t=16$. (c) $t=32$. (d) $t=64$. (e) $t=128$. (f) $t=256$.
}
\label{landfig}
\end{figure}

\begin{figure}
\centering
\centerline{
    \epsfxsize=8cm
    \epsffile{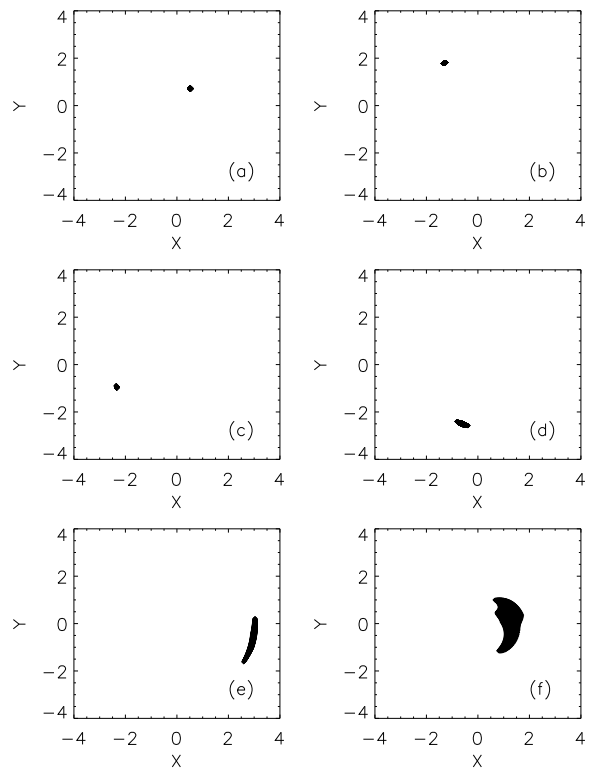}
      }
    \begin{minipage}{10cm}
    \end{minipage}
    \vskip -0.3in\hskip -0.0in
\caption{The same as the preceding Figure, now allowing for an oscillatory
perturbation of the form given by eq.~(6) with $M_{0}=0.5$, $p=2$, and
${\omega}=0.035$.
}
\label{landfig}
\end{figure}

That it may be possible to achieve efficient chaotic phase mixing in an
oscillating galactic potential while still relaxing towards a nearly integrable
state within $10-20t_{D}$ or so is illustrated visually in Figures 10 - 12,
which exhibit the behaviour of the same localised ensemble of 1600 initial
conditions, again subjected to damped oscillations of the form given by eq.~(6)
with $m_{0}=0.5$, $p=2$, and $t_{0}=100$ for three different choices of 
frequency ${\omega}$. Figure 9 exhibits the results of a corresponding
evolution in the presence of a nonoscillatory perturbation of the form given
by eq.~(8). In each case, the six panels exhibit the $x$ and $y$
coordinates of each of the orbits at times $t=0$, $16$, $32$, $64$, $128$,
and $256$, the last corresponding to an interval of approximately $12.8t_{D}$.

It is evident visually from Figure 9 that, in the absence of oscillations,
mixing is comparatively inefficient. Indeed, the evolution in that Figure is
qualitatively identical to examples of regular phase mixing in time-independent
potentials ({\em cf.} Figure 2 in Kandrup 1999). As asserted in Section 3,
a non-oscillatory perturbation of the integrable Plummer potential leads to
little if any transient chaos and, as such, no evidence for chaotic phase
mixing. The remaining three Figures, which incorporate a systematic pulsation, 
all yield phase mixing that is substantially more efficient. Figure 10, which 
represents orbits that have been pulsed with ${\omega}=0.035$, a frequency 
near the lower edge of the resonance, clearly exhibits more robust mixing, 
although the mixing is still considerably less efficient than what is observed 
for strongly chaotic flows in time-independent potentials ({\em cf.} Figure 1 
in Kandrup 1999). Indeed, it is difficult to determine unambiguously whether
the orbits used to generate this Figure genuinely exhibit significant transient
chaos.

By contrast, Figure 11, which was generated for ${\omega}=0.70$, exhibits
precisely the sort of behaviour which one has come to associate with chaotic
phase mixing in time-independent potentials. For the first two dynamical
times, the localised ensemble of orbits, which started with a phase space 
size $<10^{-2}$, still remains confined. By $t=64$, however, corresponding
to an interval ${\approx}{\;}3.2t_{D}$, the orbits have begun to spread 
significantly; and, by $t=128$, corresponding to roughly $6.4t_{D}$,
the ensemble has dispersed to fill a large portion of the available phase
space. Nevertheless, despite this efficient mixing, the potential is damping
rapidly towards a near integrable state. By a time $t=256$, the overall 
amplitude of the perturbation $m_{0}/(1+t/t_{0})^{2}$, has decreased from 
$m_{0}=0.5$ to $m_{0}{\;}{\approx}{\;}0.04$, this corresponding to a much more
nearly time-independent system.

Figure 12 exhibits an example of `incomplete' chaotic phase mixing for the
case of a frequency ${\omega}=3.5$. Here again one sees clear evidence
of the dispersive behaviour indicative of chaotic phase mixing; but, since
one is closer to the upper edge of the resonance, the efficient mixing has 
{\em de facto} turned off before the ensemble could fill all the accessible
phase space. The dearth of orbits near $x=y=0$ is exactly what one would
expect in a time-independent spherical potential for orbits with an 
appreciable angular momentum.

\begin{figure}
\centering
\centerline{
    \epsfxsize=8cm
    \epsffile{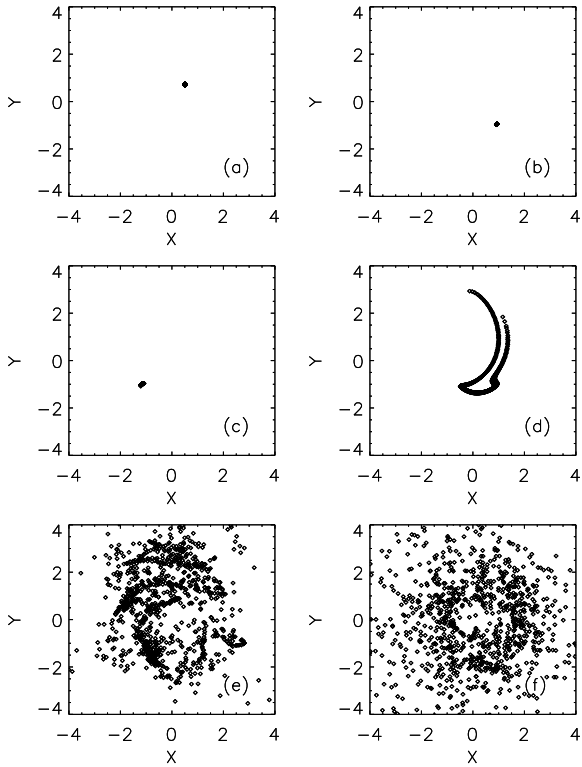}
      }
    \begin{minipage}{10cm}
    \end{minipage}
    \vskip -0.3in\hskip -0.0in
\caption{The same as the preceding Figure, for ${\omega}=0.70$.
}
\label{landfig}
\end{figure}

\begin{figure}
\centering
\centerline{
    \epsfxsize=8cm
    \epsffile{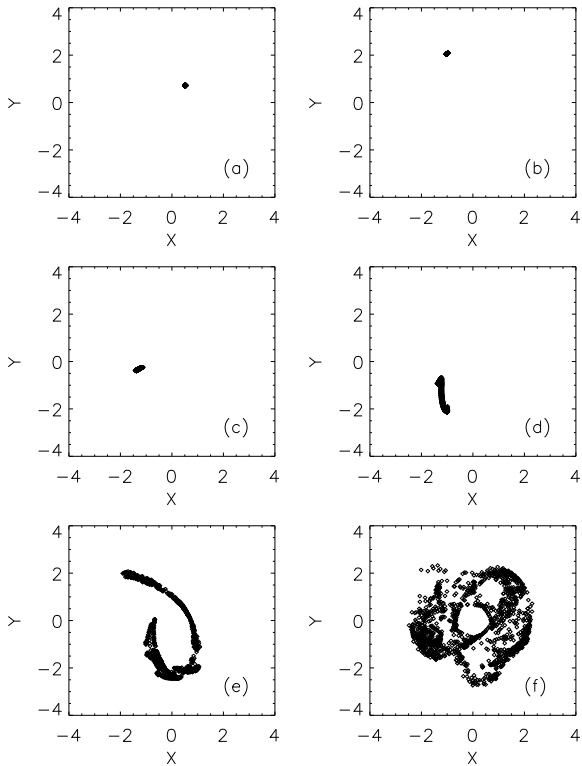}
      }
    \begin{minipage}{10cm}
    \end{minipage}
    \vskip -0.3in\hskip -0.0in
\caption{The same as the preceding Figure, for ${\omega}=3.50$.
}
\label{landfig}
\end{figure}

\section{DISCUSSION}
The numerical computations described herein lead to several seemingly
unambiguous conclusions:
\par\noindent
1. Subjecting orbits to a possibly damped oscillatory time-dependence can 
give rise to substantial amounts of transient chaos, even if the initial
and final form of the potential is completely integrable.
\par\noindent
2. This transient chaos appears to arise from a resonant coupling which,
at least for large amplitudes, is very broad and, as such, might be expected
to be present in many physical systems.
\par\noindent
3. This transient chaos can drive chaotic phase mixing which, in the context
of a fully self-consistent evolution, might be expected to play an important 
role in violent relaxation, {\em e.g.}, during the formation of galaxies and 
galaxy halos and in mergers/close encounters of galaxies.

Three extensions of this work are underway: The first involves a consideration 
of additional examples with the aim of determining the extent to which this 
transient chaos
is generic. The examples described here suggest that transient chaos
can arise whenever the perturbation has substantial power at frequencies
that are comparable to within an order of magnitude or so to the natural
frequencies associated with the orbits. However, this needs to be checked 
in the context of other, more realistic models.

A second extension involves efforts to better understand the origins of this
chaos using geometric arguments due originally to Pettini (1993). Recently,
Kandrup, Sideris, \& Bohn (2002) showed that thermodynamic arguments
proposed originally to predict the size of the largest Lyapunov exponent in
high-dimensional time-independent Hamiltonian systems (Casetti, Clementi,
\& Pettini 1996) also work surprisingly well in lower-dimensional ({\em e.g.,} 
two- and three-degree-of-freedom) systems. The aim is to adapt that work to 
the case of Hamiltonian systems admitting a time-dependence of the form that 
one might expect to encounter in the context of violent relaxation.

The third extension involves searching for evidence of transient chaos in
the context of fully self-consistent numerical simulations, both for 
self-gravitating systems and for non-neutral plasmas, {\em e.g.,} charged
particle beams. Recent simulations ({\em e.g.,} Kishek, Bohn, Haber, O'Shea,
Reiser, \& Kandrup 2001) have shown that fully self-consistent
simulations of beams can exhibit evidence of chaotic phase mixing. The 
obvious issue is whether, as one might expect, the degree of chaotic mixing 
observed in such a beam or in a simulation of violent relaxation in a 
self-gravitating system correlates with the degree to which the bulk potential
admits a significant time-dependent oscillatory component.

Plummer spheres do not constitute especially realistic models of early-type
galaxies: aside from the idealisation of spherical symmetry, real galaxies
often have a central cusp and their densities fall off much more slowly
at large radii. Despite this, however, it is interesting to estimate more
carefully the `real' values that ${\Omega}$ and ${\omega}$ might assume if
nature did somehow construct Plummer spheres. 

To estimate a reasonable
value for the pulsation frequency one can proceed as follows:
In the dimensionless units used in this paper, the half-mass radius $r_{h}$
inside of which half the mass of the galaxy is contained satisfies 
$r_{h}{\;}{\approx}{\;}1.3$. However, ${\overline r}(E)$, the mean radius
associated with a uniform sampling of an $E=$ constant hypersurface, satisfies
${\overline r(E)}=r_{h}$ for $E{\;}{\approx}{\;}0.4$. One might thus argue
that a `typical' frequency associated with the system as a whole would 
correspond to a characteristic frequency associated of orbits with 
$E{\;}{\sim}{\;}0.4$ which, by analogy with Figure 5, can be computed as 
${\Omega}_{m}{\;}{\sim}{\;}0.4$. 
Alternatively, given that the density ${\rho}=(3/4{\pi})(1+r^{2})^{-5/2}$,
it is easily seen that the half mass density ${\rho}_{h}={\rho}(r_{h})$ 
defines a characteristic
frequency ${\Omega}_{h}=\sqrt{4{\pi}G{\rho}_{h}}{\;}{\approx}{\;}0.50$.
One might, therefore, anticipate that realistic bulk oscillations associated
with the system would be characterised by frequencies 
$0.4{\;}{\la}{\;}{\omega}{\;}{\la}{\;}1.0$, the larger values corresponding
to `higher order modes.'
Alternatively, an examination of Fourier spectra for representative orbit
ensembles at different energies shows that the peak frequency ranges from
${\Omega}_{m}\to 1.0$ for $E\to -1.0$ to, {\em e.g.,} 
${\Omega}_{m}{\;}{\sim}{\;}0.05$ for $E=-0.1$. 

To the extent that the resonance described in this paper is representative
of the Plummer sphere, it would thus seem likely that all but the very highest
energy orbits would in fact be able to resonate with a large scale
`bulk' oscillation. One might also be tempted to conclude that the possibility
of resonances for ${\omega}{\;}{\gg}{\;}{\Omega}_{m}$ is largely unimportant
physically. This, however, is not necessarily the case. One anticipates 
generically that, as the system `relaxes', power will cascade from larger 
scales and lower
frequencies to shorter scales and higher frequencies. To the extent, however,
that the natural frequency of the perturbation is more important than its
detailed form, one might expect that such higher frequency oscillations, once
triggered, could play an important role in continuing the process of violent
relaxation on shorter scales. This is, {\em e.g.,} consistent with the fact
(Kandrup {\em et al} 2003) that shorter scale perturbations associated with
supermassive black hole binaries orbiting near the center of a galaxy can
trigger qualitatively similar effects in terms of resonant phase mixing and
energy and mass transport.

\section*{Acknowledgments}
The authors acknowledge useful discussions with Bal{\v s}a Terzi{\'c}.
HEK and IVS were supported in part by NSF AST-0070809. IVS was also supported
in part by Department of Education grant G1A62056.

\vfill\eject

\end{document}